\documentstyle[amssymb,rmp,aps]{revtex}

\begin{document}
\title{Condensed phases of gases inside nanotube bundles}
\author{M. Mercedes Calbi, Milton W. Cole}
\address{Department of Physics, Pennsylvania State University, University
Park, Pennsylvania 16802, USA}
\author{Silvina M. Gatica}
\address{Departmento de F\'{\i}sica, Facultad de Ciencias Exactas y 
Naturales, \\ 
Universidad de Buenos Aires, 1428 Buenos Aires, Argentina}
\author{Mary J. Bojan}
\address{Department of Chemistry, Pennsylvania State
University, University Park, Pennsylvania 16802, USA}
\author{George Stan }
\address{Institute for Physical Science and Technology, and Department of
Chemical Engineering, \\
University of Maryland, College Park, Maryland 20742, USA}
\date{\today}
\maketitle

\begin{abstract}
An overview is presented of the various phases predicted to occur when gases
are absorbed within a bundle of carbon nanotubes. The behavior may be
characterized by an effective dimensionality, which depends on the species
and the temperature. Small molecules are strongly attracted to the
interstitial channels between tubes. There, they undergo transitions between
ordered and disordered quasi-one dimensional (1D) phases. Both small and
large molecules display 1D and/or 2D phase behavior when adsorbed within the
nanotubes, depending on the species and thermodynamic conditions. Finally,
molecules adsorbed on the external surface of the bundle exhibit 1D behavior
(striped phases), which crosses over to 2D behavior (monolayer film) and
eventually 3D behavior (thick film) as the coverage is increased. The
various phases exhibit a wide variety of thermal and other properties that
we discuss here.
\end{abstract}

\tableofcontents

\section{Introduction}

A carbon nanotube is a cylindrical tube, of typical radius between 5 and 10 
\AA , consisting of one or more concentric rolled up planes of graphite. The
discovery of such tubes by Iijima, in 1991, has elicited numerous ideas of
interest to both fundamental and applied scientists (Dresselhaus {\em et al}%
., 1996). For physicists, much of the excitement has focused on the
possibility of observing novel one-dimensional (1D) behavior, a consequence
of the remarkable aspect ratio ($\thickapprox $10$^{4}$) of the length of
the tubes to their radius. Such 1D behavior is expected for both the tubes
themselves and any gas imbibed within the tubes; the latter is the subject
of this Colloquium article. Specifically, we explore the phases of the
adsorbate present at each of the various sites of a bundle (or ``rope'') of
carbon nanotubes depicted in Fig. 1. These three regions, assumed to
contribute additively to the gas uptake, are inside a tube, within an
interstitial channel (the space between tubes, abbreviated IC) and the
external surface of the bundle. We shall see in this article that various
phases have been predicted whose properties are quite remarkable.
Experimental tests of these predictions are under way in many laboratories
(Dresselhaus {\em et al}., 1999; Teizer {\em et al}., 1999, 2000; Bienfait 
{\em et al.}, 2000; Kuznetsova {\em et al.}, 2000; Muris {\em et al.}, 
2000; Talapatra {\em et al}., 2000; Weber {\em et al}., 2000). 
Here, we focus on
qualitative aspects of the predictions without providing a detailed
explanation of the theoretical methods used to derive them. The key
assumptions used in the calculations (e.g., that the nanotube lattice is
perfectly ordered) are discussed in the final section of this article.

The word ``dimension'' should be used with some caution in describing these
phenomena. In statistical mechanics, a system is usually characterized as
``D-dimensional'' if D equals the number of spatial dimensions which diverge
when the thermodynamic limit is taken. In the following, the geometry is
fixed as that shown in Figure 1, a hexagonal array of nanotubes. The length
of the tubes is assumed to be infinite. We shall see how temperature ($T$)
influences the observed dimensionality of the adsorbate. This happens, for
example, when certain degrees of freedom are effectively frozen out at low $T
$, leaving only excitations characterized by a reduced dimensionality. This
change of effective dimension D affects the power law behavior and the
critical exponents characterizing the divergent behavior of thermodynamic
properties at phase transitions. The value of D may either increase or
decrease as a function of $T$. In two cases discussed below (single particle
and phonon excitations), D is an increasing function of $T$, while in
another (condensation transition) it is a decreasing function of $T$.

A typical adsorption experiment involves the presence of a known vapor which
is brought into contact with the nanotube bundle. An obvious question
arises: at a given $T$ and external pressure ($P$), how much gas goes into
each of the various sites? Fig. 2 presents the result of model calculations
of the uptake of various gases, assuming equilibrium between the absorbed
and the external gases. Each species of gas has an interatomic interaction $%
V(r)$ characterized by energy ($\epsilon $) and length ($\sigma $)
parameters. Typically, but not always, one assumes that this has the
Lennard-Jones form:

$\qquad $%
\begin{equation}
V(r)=4\epsilon \{[\sigma /r]^{12}-[\sigma /r]^{6}\}\qquad 
\end{equation}

One observes in Fig. 2 that species having small values of $\sigma $ (i.e.,
He, Ne, and H$_{2}$) absorb significantly within the IC's, while larger
molecules do not. The word ``significant'' is defined here to mean that the
average spacing (inverse 1D density) between molecules is less than 10 $%
\sigma $. The $\sigma $-dependent behavior seen in Fig. 2 is a logical
consequence of a key assumption used to create the figure: that the bundle
of tubes does not deform (e.g. swell) during the absorption process. Hence,
large molecules don't fit within the IC's.

Both these small and quite large (CF$_{4}$ and SF$_{6}$) molecular species
are seen in Fig. 2 to absorb strongly inside the tubes. In contrast,
intermediate size gases (CH$_{4}$, Kr and Xe) do not populate the tubes
significantly at the assumed pressure and temperature ($T^{\ast
}=k_{B}T/\epsilon =1$), although they do at somewhat lower temperature and
higher pressure. The determinant of this systematic behavior is a
competition between adhesive forces, favoring uptake, and cohesive forces,
which oppose uptake; these latter predominate at large values of $\epsilon $%
, as seen in the figure. Ar (shown as an open circle in Fig. 2) is a system
for which these competing factors are comparable; its 1D density at the
conditions of Fig. 2 is 0.1/$\sigma $, so it is a borderline case.

The adsorbate's behavior within each region of space is a function of the
number of molecules there. What we know for certain is that the total
number, $N$, of absorbed molecules increases with pressure $P$. An important
thermodynamic parameter characterizing this dependence is the chemical
potential

\begin{equation}
\mu =\left( \frac{\partial F}{\partial N}\right) _{T}\ 
\end{equation}%
where $F$ is the Helmholtz free energy. In equilibrium, the chemical
potentials of the coexisting adsorbate and vapor phases coincide. For an
ideal gas at temperature $T$, we have the relation

\begin{equation}
\beta \mu =ln(\beta P\lambda ^{3})
\end{equation}%
where $\beta \equiv 1/(k_{B}T)$ and $\lambda $ is the de Broglie thermal
wavelength, i.e., a typical particle's wavelength:

\begin{equation}
\lambda ^{2}=\frac{2\pi \beta \hbar ^{2}}{m}
\end{equation}%
$m$ being the mass of the particle \footnote{%
Eq. 3 assumes that the gas is monatomic and spinless. Otherwise, the
argument of the logarithm is to be divided by the internal partition
function of the adsorbate.}. A benchmark for the chemical potential is its
value ($\mu _{0}$) at saturated vapor pressure. Capillary condensation is
the formation of liquid within pores at $\mu <\mu _{0}$.

The task of the theory is to compute how $\mu $ of the adsorbate varies with 
$N$; using Eq. 3, one can then predict the dependence of $N$ on $P$.
Conversely, experiments (adsorption isotherms) which measure $N(P)$ yield
the chemical potential of the adsorbate. This is an important goal because
all thermodynamic variables characterizing the film can be derived from $\mu
(N,T)$, since its integral over $N$ is the free energy $F$, from which other
thermodynamic properties can be computed. More interesting, perhaps, is that
the structure of the film can often be deduced by judicious interpretation
of the $\mu $ data. Fig. 3 exemplifies this; the $T=0$ isotherm for H$_{2}$
is seen to exhibit a series of steps, due to a succession of absorption
phenomena: covering the nanotube interior walls (``shell'' phase), filling
the groove on the outside of the bundle, filling the IC's, filling the axial
region of the tubes, and eventually fully covering the outside of the bundle
at higher chemical potential than is shown in the figure. The size of each
step is equal to the number of molecules ``filling'' each site. This site
capacity, per unit length of bundle, is equal to a product of the 1D density
within the site and the number of equivalent sites within the bundle. If we
let $j$ equal the number of tubes on each facet bounding the bundle, assumed
to be hexagonal ($j=4$ in Fig. 1), then one may prove that the number of
external grooves is $6(j-1)$, the number of IC's is $6[j(j-2)+1]$ and the
number of tubes is $1+3j(j-1)$. For large $j$, there are twice as many IC's
as tubes. We note that the steps seen in Fig. 3 become progressively rounded
as $T$ increases.

One of the practical motivations for studying adsorption in nanotubes is
their potential for high storage capacity space. While not the focus of this
article it is worthwhile to discuss this topic briefly. There are several
encouraging aspects to this pursuit. One is that graphitic materials provide
a particularly strong atraction as compared with another adsorbents (Bruch 
{\em et al}., 1997). A second is that curved surfaces provide relatively
attractive binding environments compared to planar surfaces. Finally, and
most significantly, single wall carbon nanotubes are optimally efficient 
adsorbents on a ``per gram'' basis. The reason quite simple is that they are
``all surface''(Williams and Eklund, 2000; Dillon and Heben, 2001). 
For example, in Fig. 3, the uptake at quite low chemical
potential (far below saturation) exceeds 10 mmole/gr. A typical ``high
surface area'' material, such as grafoil, adsorbs $\thickapprox $ 0.5
mmole/gr under these conditions. Thus the nanotubes provide an order of
magnitude increase in storage capacity.

In the following sections, we describe results of calculations of phase
behavior and other properties within the various regions of space where the
particles are absorbed. We ignore almost entirely the methodology of the
calculations; the interested reader may turn to the original sources. We
focus on predictions of the phase behavior since experiments have not yet
tested these predictions.

\section{Gases within interstitial channels}

As indicated in Fig. 2, we anticipate that He, Ne and H$_{2}$ will adsorb
within the IC's in large numbers. What properties do we expect to observe?
The molecules are confined to the vicinity of the z-axis, defined to lie at
the center of a given channel, and thus 1D behavior is predicted (Stan 
{\em et al}., 1998). Fig. 4
depicts the potential energy experienced by a H$_{2}$ molecule as it moves
away from the axis and the corresponding wave function, which extends only $%
\thickapprox $ 0.2 \AA\ from this axis. Such a high degree of confinement
suggests a 1D interpretation of the behavior. Interestingly, two extreme
alternative 1 D models have been employed to characterize this system. A
``quasi-free'' model neglects any z dependence of the potential energy. A
``localized'' model (which has been justified by a band structure
calculation for one specific array of tubes (Cole {\em et al}., 2000))
assumes instead that the carbon environment provides a set of 1D periodic
sites into which the particles settle. Each model's simplification permits
direct contact with a body of previous theoretical work on 1D models or
permits one to solve a new and relatively simple 1D problem (Takahashi,
1999). Particularly interesting is the fact that the localized description
represents a realization of the famous lattice gas model in statistical
mechanics. In this model, sites may have occupancy zero or one; in the
simplest case, only nearest neighbours interact. This model is often applied
(perhaps surprisingly) to cases, such as the liquid-vapor transition in free
space, for which the model is a significant abstraction of reality. In such
cases, the model's limitations are known and attention focuses on the
critical regime, for which the essential physics does not suffer from the
model's simplifications.

Remarkably, many of the salient features of the collective behavior
predicted by the two contradictory models are similar. Both models have the
feature that no phase transition occurs at $T\neq 0$ for a system consisting
of a single IC; this is because thermal fluctuations disrupt any
hypothetical order in 1D. At $T=0$, however, transitions for a single IC may
occur as a function of density. For example, within the quasifree 1D model, H%
$_{2}$ undergoes both a condensation and a freezing transition if the free
space interaction is assumed (Gordillo {\em et al}., 2000); if screening is
taken into account (Kostov {\em et al}., 2000), however, the condensation
transition is suppressed and the single IC system's ground state is a gas!
This remarkable behavior is of great interest, of course, falling into the
category of so-called ``quantum phase transitions''(Sachdev, 1999).

The interactions between molecules in adjacent IC's is essential to the
formation of ordered phases at finite $T$. Because the separation between
IC's is large ($d\thickapprox 10$\AA ), this interchannel interaction is
quite weak, smaller than the pair well depth by a factor of order $(\sigma
/d)^{6}$, where $\sigma $ is the molecular diameter. Hence the transverse
interaction affects behavior only at low $T$, but the consequence can be
dramatic: phase transitions occur which do not occur without these
interactions. Figure 5 depicts such an anisotropic condensed phase which is
expected to occur at low $T$. Carraro has studied the properties of
anisotropic crystalline phases arising from the interchannel interaction
(Carraro , 2000). Because this interaction is weak, that crystal melts at
very low $T$ to a liquid phase, which remains condensed up to a higher
temperature. A quantum system (one with weak interactions and light mass,
such as He) will not exhibit any such crystalline phase in the absence of an
ordering external potential (as is also the case for He in 3D unless
pressure is applied), but a liquid phase should occur.

Figure 6 exemplifies the He system's properties, studied within a
(localized) anisotropic Ising model. In this model, atoms reside at sites
which are closely spaced along the z axis and interact with one another as
well as with atoms on adjacent chains (Cole {\em et al}., 2000). The
interaction perpendicular to the IC's is $\thickapprox $ 1 \% of that along
the axes. The heat capacity is seen to correspond closely to that of a
strictly 1D system (a smooth curve) down to $k_{B}T\thickapprox J_{z}$ , the
nearest neighbor interaction. While the strictly 1D system exhibits a gentle
maximum in this region, near $T\thickapprox 0.8 J_{z}/k_{B}$, a phase
transition occurs in the system with interacting channels. This phase
transition is a condensation of the anisotropic fluid, with droplets
extending into contiguous channels; indeed they extend a distance equal to
the correlation length, which diverges at the critical temperature, $T_{c}$%
(see Fig. 6). The value of $T_{c}$ is high compared to the interchannel
interaction temperature ($\thickapprox 0.015 J_{z}/k_{B}$). This follows from
the fact that the transition involves a large block of molecules (those
within a 1D correlation length) within one IC interacting with a similar
block in an adjacent IC. An analytic formula derived by Fisher (1967)
provides the value of $T_{c}$ in the limit of strong anisotropy, as is the
case here. One of the most interesting features of the transition seen in
figure 6 is that the heat capacity attributable to the absorbate actually
dwarfs that of the host material. The reason is simply that the nanotubes,
consisting of strongly cohering carbon, with a Debye temperature $%
\thickapprox 2000K$, have by themselves only a negligible heat capacity at
low $T$ (Mizel {\em et al}., 1999).

Figure 4 illustrates another, rather dramatic effect. Because H$_{2}$
molecules are tightly confined within the IC's, their zero point energy of
motion perpendicular to the z axis is large. A tiny (0.5\%) increase in the
spacing between tubes greatly reduces this energy. The net result, including
the energy cost of spreading the tubes, is a doubling of the binding energy
within the IC's (Calbi {\em et al}., 2000). This is a cooperative effect,
requiring the presence of many molecules in order to obviate bending of the
tubes. The resulting expanded lattice represents the ground state of the
system. This expansion of the lattice is large enough to be detected in
diffraction (Amelinckx {\em et al}., 1999) or Raman measurements of the
tubes' breathing modes (Venkateswaran {\em et al}., 1999; Dresselhaus and
Eklund, 2000). This sensitivity of the thermodynamic and structural
properties to the zero point energy has led Johnson and coworkers to propose
using nanotube bundles to separate isotopes of hydrogen (Wang {\em et al}.,
1999).

\section{Matter within the tubes}

\subsection{Cylindrical model}

We consider next the behavior of molecules adsorbed within the nanotubes
themselves. We begin with the single particle problem of evaluating the
potential energy and the corresponding wave functions of the Schr\"{o}dinger
equation. This is a cylindrical surface variant of the familiar ``particle
in a box'' problem. The square of the wave function is the probability
density for the particle. As Fig. 7 exemplifies, a molecule with a diameter $%
\sigma $ small compared to the tube diameter is bound strongly to the
vicinity of the tube wall. This localization in the radial direction
justifies a simple ``cylindrical model'' of the film's properties.
Specifically, this model assumes that the particles are confined to the 2D
cylindrical surface, $r=R_{eq}$, the equilibrium distance of the particle
from the axis\footnote{%
As exemplified in Fig. 7, the anharmonicity of the potential may give rise
to a significant displacement of the position of maximum probability
relative to the potential energy minimum.}. In the classical case, such
confinement occurs at low temperature and is described by the Boltzmann
factor in the single particle density distribution ($\backsim \exp [-\beta
V(r)]$). In the extreme quantum case of He or H$_{2}$, the ground state
density is spread over a range $\Delta r\thickapprox $ 0.2 \AA\ due to the
zero-point motion in this potential, as seen in Fig. 7. Even in that case,
however, the cylindrical model is useful because the radial degree of
freedom is {\em de facto} frozen out due to the high energy scale associated
with radial excitation ($\thickapprox 100$ K).

Determining the quantum states of the system is easy if we ignore all 
dependence of
the adsorption potential on the variables $z$ and $\varphi $, the azimuthal
angle (Stan and Cole, 1998). With these assumptions, each energy level of a
single particle moving on the surface $r=R_{eq}$ is a sum of a constant (the
radial ground state energy $E_{r}$), plus azimuthal and axial kinetic
energies:

\begin{equation}
E=E_{r}+E_{\varphi }+E_{z}
\end{equation}
where

\begin{equation}
E_\varphi =\frac{(\hbar \nu )^{2}}{2mR_{eq}^{2}}
\end{equation}
and

\begin{equation}
E_{z}=\frac{(\hbar k_{z})^{2}}{2m}
\end{equation}

Here, the angular momentum about the $z$ axis is $\nu \hbar $, where $\nu $
is zero or an integer. Thus the azimuthal motion is characterized by a
discrete spectrum; the gap between the two lowest levels ($\nu $=0 and $\pm $%
1) corresponds to an azimuthal excitation temperature, $T_{\varphi }=(\hbar
\nu )^{2}/(2mR_{eq}^{2}k_{B})$, which is of order 1 K for He and H$_{2}$ and
0.1 K for Ne. This temperature divides the behavior into a low temperature
regime, for which $\nu =0$, and a high temperature regime, where a
significant range of $\nu $ states is populated. For $T<<T_{\varphi }$, the
only degree of freedom which is excited is that associated with $E_{z}$, the
quasicontinuous energy of motion parallel to the axis. This regime therefore
corresponds to 1D free particle motion. From the analogous problem in 3D, we
anticipate the result that the specific heat $c(T)$ (per particle)
approaches the 1D limit $c\rightarrow k_{B}/2$ at low temperature . One sees
this behavior in Fig. 8; near $T_{\varphi }$ the specific heat evolves
nonmonotonically \footnote{%
The nonmonotonic behavior in Fig. 8 is similar to that found for $c(T)$ of
the linear rotor, for which the eigenvalues vary as $\nu (\nu +1),\nu
=0,1,...$The more familiar monotonic behavior occurs in cases when the
energy level spacing is relatively constant.} from the 1D value to the 2D
value, $c=k_{B}$ (occurring at high $T$ due to excitation of azimuthal
motion). This dimensional crossover in the quantum case has no classical
analogue because it is a consequence of the quantization of azimuthal
motion, a wave property\footnote{In contrast, a reduction in dimensionality 
(from 3D to 2D) associated with particles coalescing onto the cylindrical 
surface $r=R_{eq}$ occurs for both
classical and quantum particles at high temperature.}$^,$\footnote{%
Note that Fig. 8 shows the classical specific heat, ignoring quantum
statistics as well as interactions.}.

\subsection{Collective properties}

Figure 7 hints that particles can reside in either of two regions of space-
near the wall or near the axis. We first consider adsorption in a dense
``cylindrical shell'' phase, localized near the wall. The collective
behavior of molecules in this phase is analogous to that within a monolayer
film in several respects, but intriguing differences should occur. The
specific behavior is sensitive to the species being adsorbed. In some cases,
the atomicity of the nanotubes gives rise to a periodic potential
(``corrugation'') which leads to commensurate phases of the adsorbate, while
in other cases, that corrugation can be neglected. This variable behavior is
analogous to that on graphite (Bruch {\em et al}., 1997), where commensurate
phases dominate the monolayer phase diagrams of some adsorbates, like He and
H$_{2}$, but not others (e.g. Ar). One qualitative difference is that the
nanotubes' cylindrical shape yields a periodicity condition that encourages
a specific commensurate phase associated with this periodicity (Green and
Chamon , 2000).

One of the fundamental questions about the shell phase pertains to the
existence and nature of a (finite $T$) crystalline phase within the tubes
comprising a bundle. Recall that a 2D solid is quite different from a 3D
solid; the 2D solid is sensitive to long wavelength fluctuations that
prohibit the usual crystalline order. Hence, no delta function (Bragg) peaks
occur in the X-ray scattering from 2D solids, assuming that no periodic
potential is present. Instead, orientational long range order is permitted
in 2D, manifested by a quite distinct behavior of the correlation functions,
thermodynamic properties and transverse sound speed; near melting, for
example, the specific heat exhibits an essential singularity rather than the
discontinuities characteristic of melting in 3D (Strandburg, 1988; Glaser
and Clarke, 1993). Since a single nanotube is a 1D system, strictly
speaking, we suspect that a hypothetical solid in the tubes is even more
sensitive to such fluctuations. The very existence and, certainly, the
melting of such a solid are, thus far, unresolved subjects.

For the moment we set aside such questions and focus on manifestations of
this geometry which appear to be robust predictions of current theories of
this system. One involves the dynamics, i.e., the phonons, of the adsorbate;
for specificity, we focus on the long wavelength and low energy regime. Each
phonon mode possesses an azimuthal quantum number, $\nu =0,\pm 1,\pm 2$%
,\ldots , indicating the angular variation of the motion, and a
quasicontinuous wave vector ($q_{z}$) characterizing its z variation. The
lowest energy modes will have no $\varphi $ variation and thus the
excitation spectrum will be 1D in character, associated with the variable $%
q_{z}$. The corresponding low $T$ specific heat will be linear in $T$ (by
analogy with the 3D result, which varies as $T^{3}$). Vidales et al (1998)
have shown how this behavior crosses over to a 2D regime ($T^{2}$ 
dependence) at high 
$T$. Not surprisingly, the crossover temperature $T_{\varphi }$ can be
accurately estimated with a length-matching criterion: at $T_{\varphi }$,
the thermal phonon wavelength equals the circumference of the cylindrical
surface containing the atoms: $2\pi R_{shell}\thickapprox \hbar c/k_{B}T$,
where $c$ is the speed of sound within the cylindrical shell of radius $%
R_{shell}$.

Note that this increase of effective dimension (1D $\rightarrow $\ 2D) as $T$
increases is a common feature of the single particle behavior, discussed
earlier, and the phonon behavior. This is not always the case; indeed, the
reverse (a decrease of dimensionality) occurs in the previous section's
example of a 3D phase transition (because of interactions between particles
in adjacent IC's) evolving into 1D behavior at high $T$ (Gelb {\em et al}.,
1999).

Arguably more unusual than this shell phase is a so-called ``axial phase'',
i.e., molecules confined to the vicinity of the tube's axis (Gatica {\em et
al}., 2000). The appearance of this phase is exemplified in figures 9(a) and
9(b), the results of path integral Monte Carlo calculations of H$_{2}$
absorption. The axial phase appears as a rapid increase of density as a
function of chemical potential. In some respects, its appearance resembles
the layering transition on planar surfaces, but the latter corresponds to a
2D system while the axial phase is 1D. From another point of view, the
appearance of the axial phase is closer to the capillary condensation
transition exhibited in Figure 10 (Steele and Bojan, 1998; Gelb {\em et al}%
., 1999). While the detailed properties of the axial phase have not been
explored, for quantum systems there is the hope that this phase provides a
realization of the 1D Luttinger liquid model. That model's novel features
include the fact that the low-lying excitation spectrum of even a weakly
interacting fermi system (e.g. $^{3}$He) is characterized by excitations of
bose type, i.e. phonons (instead of the usual quasiparticles of Fermi liquid
theory). The key qualitative difference between axial phases of $^{3}$He 
and $^{4}$He
would then be the presence of spin waves in the $^{3}$He case.

\section{External surface of the nanotube bundle}

The external surface of the bundle provides an attractive potential for
molecules of all sizes and adsorption will occur there even if the tubes or
IC's are closed (Dresselhaus {\em et al}., 1999; Gatica {\em et al}., 2001).
If, instead, they are open, the fraction of particles residing on the
external surface depends on the surface to volume ratio, which varies
essentially as the inverse of the bundle radius. For a bundle of $\
\thickapprox $ 50 tubes, as in Fig. 1, roughly half of the available surface
lies on the outer surface. The groove formed by two adjacent tubes at the
surface is a particularly favorable environment because the well depth there
equals nearly twice that of a single graphene sheet, assuming additivity of
the interactions (Talapatra {\em et al}., 2000). Fig. 11 depicts the
potential energy in the case of an Ar atom. One observes a localized deep
well, which is the most favored adsorption site in this environment.

Simulation studies have revealed the occurrence of diverse phases on this
surface, as seen in Figure 12. The initial adsorption occurs in the groove,
itself; this is a 1D phase, with properties that can be computed from an
analytic 1D classical equation of state. At higher pressure, the coverage jumps
discontinuously (by a factor of three) to a striped phase, due to the 
appearance of two
new lines of particles parallel to the groove. As the chemical potential
increases, a sequence of additional jumps occurs. One represents the
formation of a full layer of stripes and the next corresponds to a bilayer's
formation. Beyond that point, subsequent growth is qualitatively similar to
that of wetting films on the surface of graphite. The transitions seen as
vertical risers in Fig. 12 are genuine thermodynamic transitions because the
assumed geometry is an infinite plane of perfectly parallel tubes. Surely,
there will occur rounding of these discontinuities for any real bundle of
tubes. Nevertheless, the real-world phenomena ought to bear a qualitative
resemblance to those found in the simulations. The detailed properties of
these phases remain to be explored.

\section{Summary}

In this Colloquium article we have chosen a few phenomena to illustrate the
rich physics present in this unusual geometry. Every aspect of the problem
has stimulated its own branch of the theory ``tree''; our selection of
topics reflects our bias. We encourage newcomers to this field to find their
own branch and explore territory as fertile as what we have discussed here.

The exotic scenarios described in this article are based on a sequence of
simplifying approximations, each of which merits scrutiny. For example, the
adsorption potentials have been derived by simply summing empirical
interactions between the adatoms and the individual carbons comprising the
tubes. Although significant quantitative uncertainty results from this
neglect of many-body effects (Kostov {\em et al}., 2000), we are
optimistically inclined to regard many of the qualitative predictions as
robust. A more serious concern, in our opinion, is the assumption that the
nanotubes are chemically and structurally perfect, forming a periodic
array. This is not at all the case for currently available samples. We
believe that the possibility of observing the zoo of phase transition
phenomena discussed here is a strong motivation for improving sample
quality. On the other hand, the effects of disorder on phase transitions is
a very important subject. Hence, experimentalists are urged to explore the
phenomena as a function of sample preparation; we are confident that
theorists will be delighted to see their data! 

\vspace{1cm}

{\bf Acknowledgements}

\bigskip 

Our work has been stimulated by discussions with many colleagues and workers
in this field. We are particularly grateful to Jayanth Banavar, Jordi
Boronat, Michel Bienfait, Massimo Boninsegni, Carlo Carraro, Moses Chan, Vin
Crespi, Peter Eklund, Carmen Gordillo, Bob Hallock, Karl Johnson, Susana
Hernandez, Milen Kostov, Aldo Migone, Ari Mizel, Dave Narehood, Paul Sokol,
Bill Steele, Flavio Toigo and Keith Williams. This research has been
supported by the Army Research Office, Fundaci\'on Antorchas, CONICET, 
the National Science Foundation and the Petroleum Research Fund
of the American Chemical Society.

\newpage

FIGURE CAPTIONS

Fig. 1. Schematic depiction of a bundle of 37 nanotubes, of radius 7 \AA .
This bundle contains 54 interstitial channels and 18 ``grooves'' on the
external surface.

Fig. 2. Location (within IC and/or tube) of absorbed atoms or molecules
within a bundle of tubes of radius 8 \AA , as a function of the
Lennard-Jones parameters which characterize the interactions between
adsorbates. Dots correspond to a sequence of increasing values of $\sigma $,
as follows: He, Ne, H$_{2}$ , Ar (open circle), CH$_{4}$, Kr, Xe, CF$_{4}$
and SF$_{6}$. Molecular species with strongly cohesive interactions (large $%
\epsilon ,$ i.e. CH$_{4}$, Kr, Xe) do not absorb at either site at the
assumed thermodynamic conditions: reduced temperature $T^{\ast }=1$ and
reduced chemical potential (relative to saturation) $\Delta \mu ^{\ast
}=(\mu -\mu _{0})/\epsilon =-10$ (Stan {\em et al}., 2000).

Fig. 3. H$_{2}$ absorption (per unit length of nanotube) as a function of
chemical potential $\mu $. The indicated plateau regions correspond to the
progressive filling of various distinct domains within the nanotube bundle 
(shown in Fig. 1). At the highest chemical potential shown, the mass of H$_2$ 
is 2.6 \% of the total mass of the system (tubes plus gas).
Derived from data of Stan {\em et al} (2000) and Calbi {\em et al} (2000).

Fig. 4. Potential energy (lower panel) and probability density (upper panel)
as a function of radial coordinate for H$_{2}$ molecules in an interstitial
channel corresponding to nanotubes of radius 6.9 \AA\ (dashed curves). Full
curves show results in the case of a 0.5 \% dilated lattice of nanotubes.
The binding energy (horizontal dotted line) is twice as large when the
lattice is dilated (Calbi {\em et al}., 2000).

Fig. 5. Schematic depiction of an anisotropic droplet of condensed adsorbate
within IC's. The loop represents the boundary of the droplet. This phase
results from the interaction between molecules in adjacent IC's.

Fig. 6. Specific heat at critical density (full curve) associated with a
condensation of atoms, due to their mutual interactions, within an array of
parallel IC's as a function of $T$ relative to the interaction energy along 
the axis ($\varepsilon \equiv 4 J_z$). Monte Carlo calculations (points) 
suggest a divergence, known to
be present in the exact calculation. Dashed curve is the nonsingular
specific heat resulting when interactions between atoms in neighboring
channels are neglected (i.e., the 1D Ising model). Broad full curve at 
bottom is the lattice specific heat (multiplied by 1000) of a pristine 
bundle of nanotubes (Cole {\em et al}., 2000).

Fig. 7. Bottom panel shows potential energy and lowest energy level
(horizontal line) of a He atom in a nanotube of radius $R=7$ \AA . Shown in
upper panel is the ground state probability density $\vert$f$_{0}\vert^{2}$. 

Fig. 8. Specific heat as a function of $T$ of an ultralow density He gas
inside a nanotube of radius 5 \AA\ depicting crossover from 1D behavior at
low $T$ to 2D behavior at high $T$. From Stan and Cole, 1998.

Fig. 9. (a) 3D density of H$_{2}$ molecules at $T=10$ K, as a function of
radial distance, inside a nanotube of radius 6 \AA\ for three different
values of chemical potential $\mu .$ (b) 2D density $\theta $ (squares, left
scale) of H$_{2}$ in shell phase per unit area of tube wall, and 1D density
(circles, right scale) of axial phase as a function of chemical potential.
From Gatica {\em et al}., 2000.

Fig. 10. Density of Kr atoms as a function of distance from the pore axis in
a weakly adsorbing cylindrical pore of radius 10 \AA . The solid curve is
the density distribution at a pressure just below the capillary condensation
pressure; the dashed curve corresponds to the filled pore just after
condensation (Steele and Bojan, 1998).

Fig. 11. Potential energy of an Ar atom in the groove. The contours
correspond to constant values $V/\epsilon_{ArC}= -25, -20, -15, -10, -5, -1$%
, from darker to lighter. The dashed lines correspond to the cylindrical
nanotube surfaces.

Fig. 12. Phase transitions occurring within Kr atoms adsorbed on the
external surface of the nanotube bundle, manifested as discontinuities in
coverage (expressed in adsorbate mmoles per gram of carbon) as a function 
of pressure. The stable values of coverage correspond
to striped phases, lying parallel to the grooves between external tubes.
From Gatica {\em et al}., 2001.

\bigskip

\bigskip

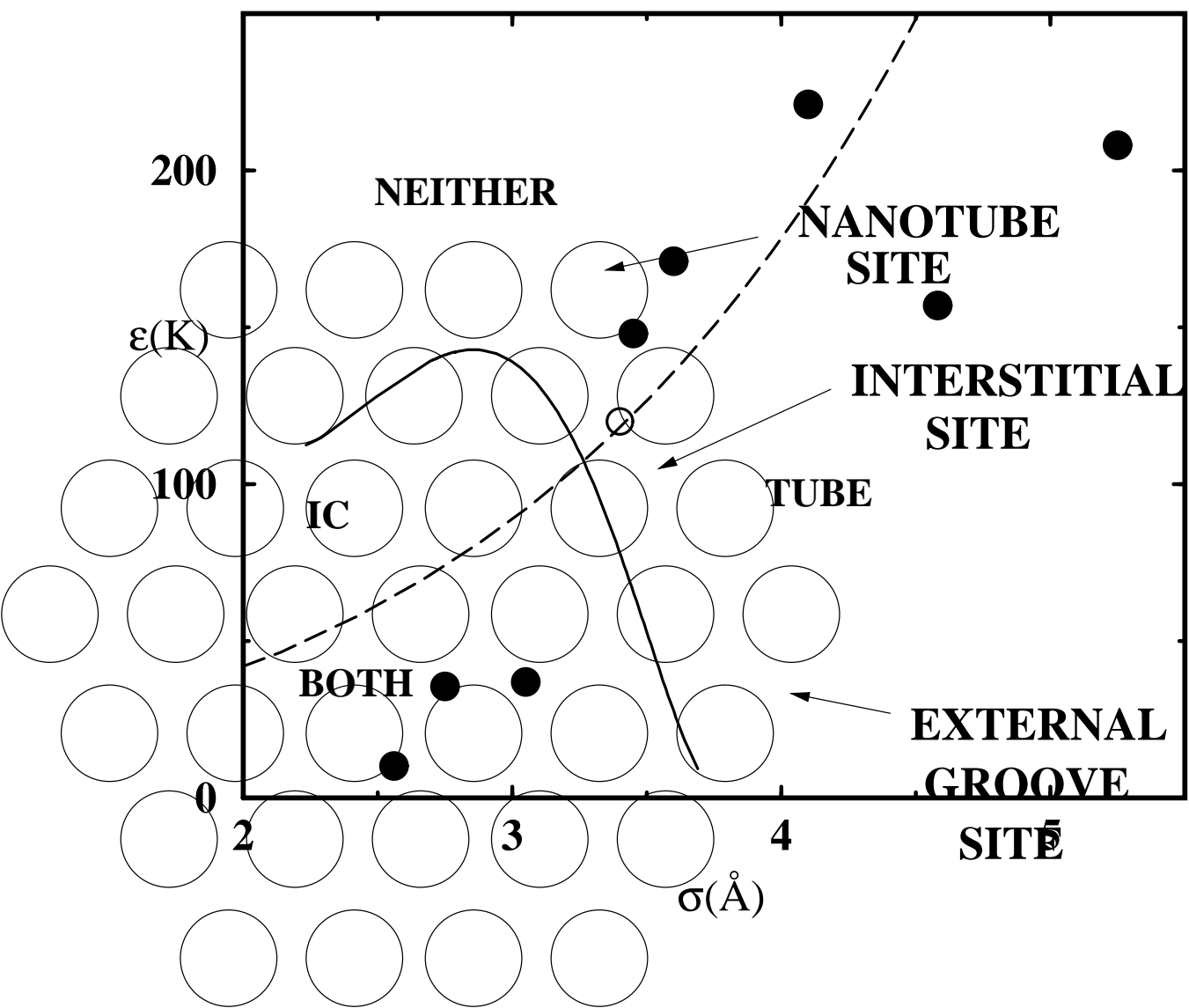
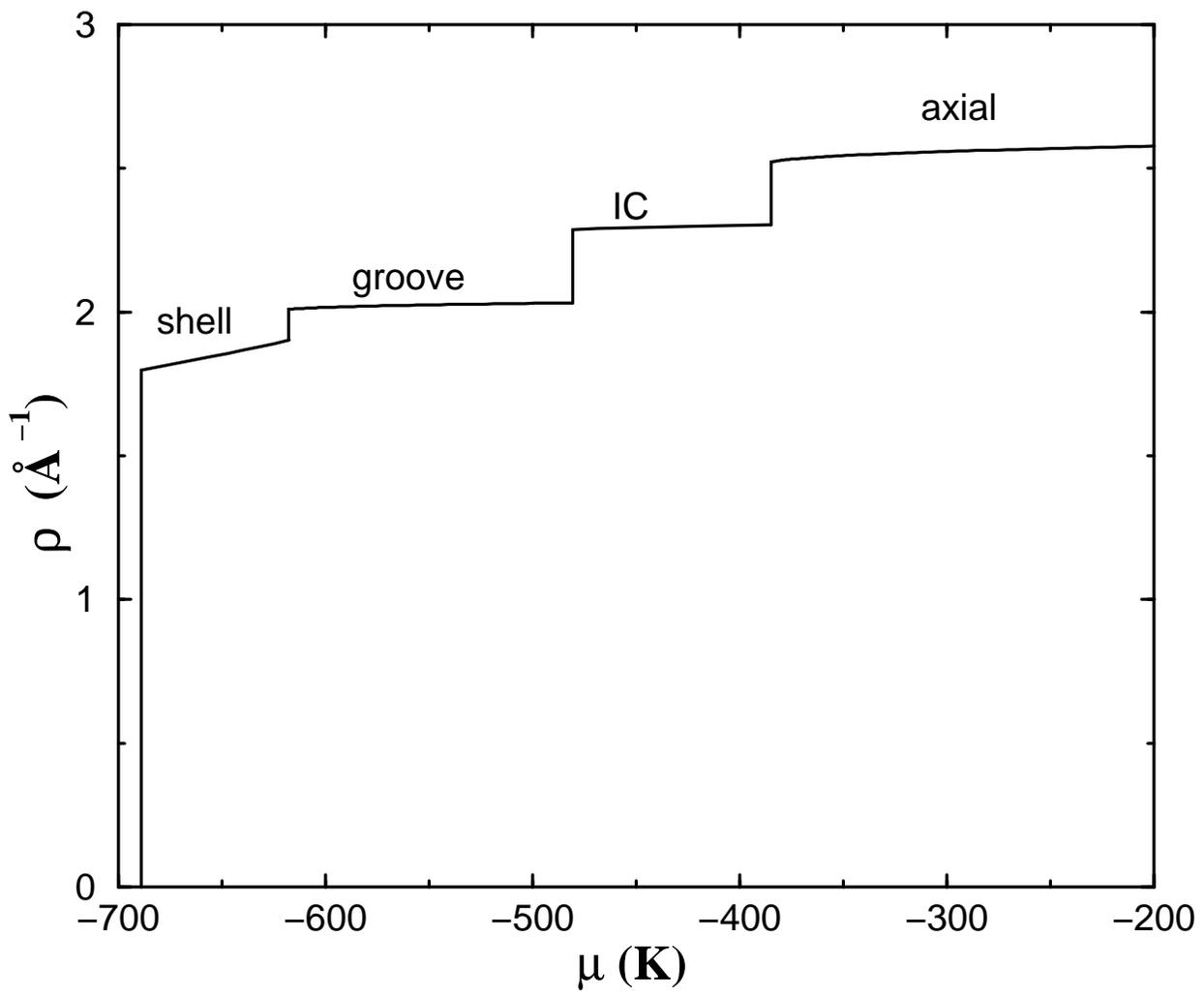
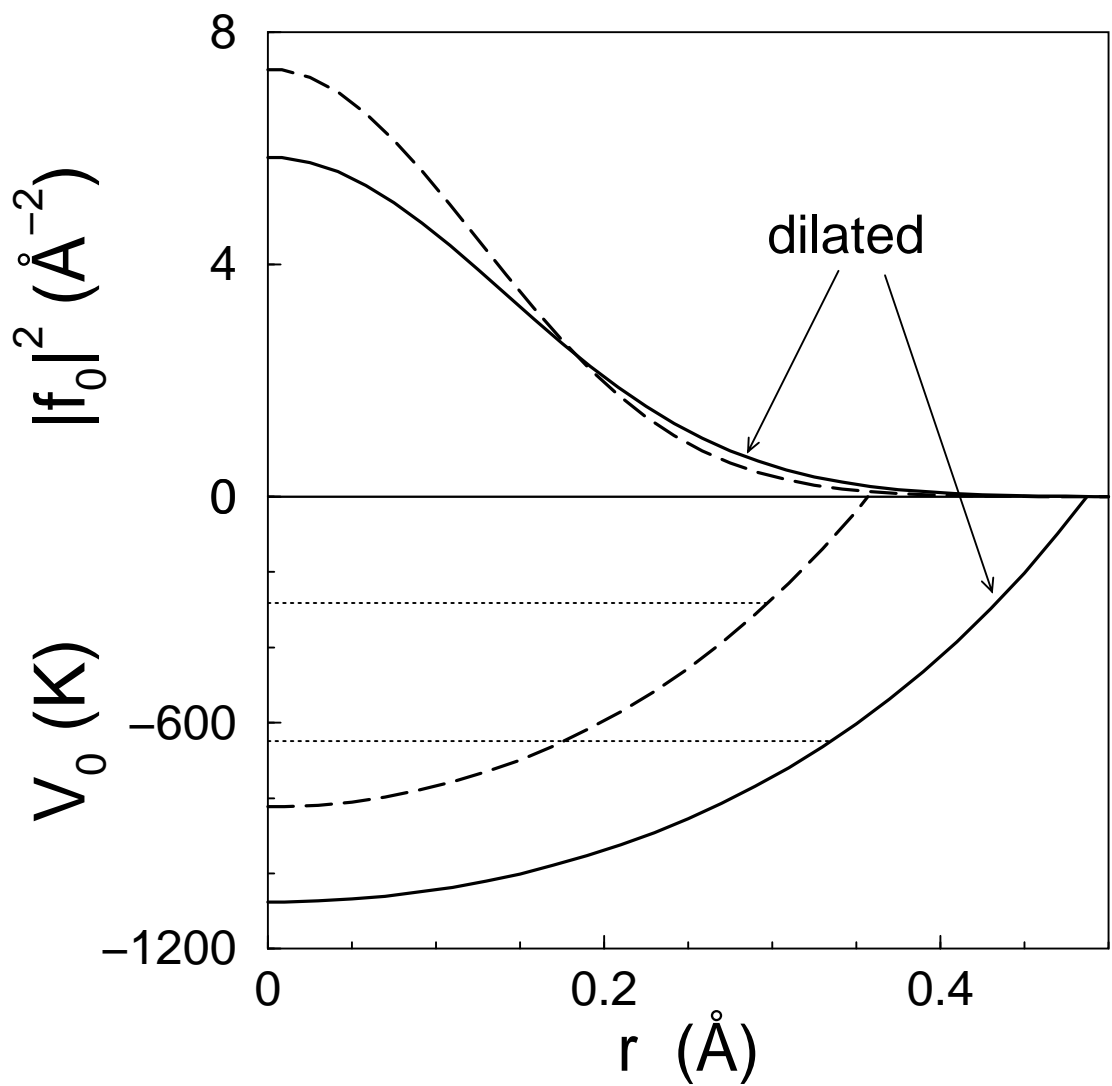
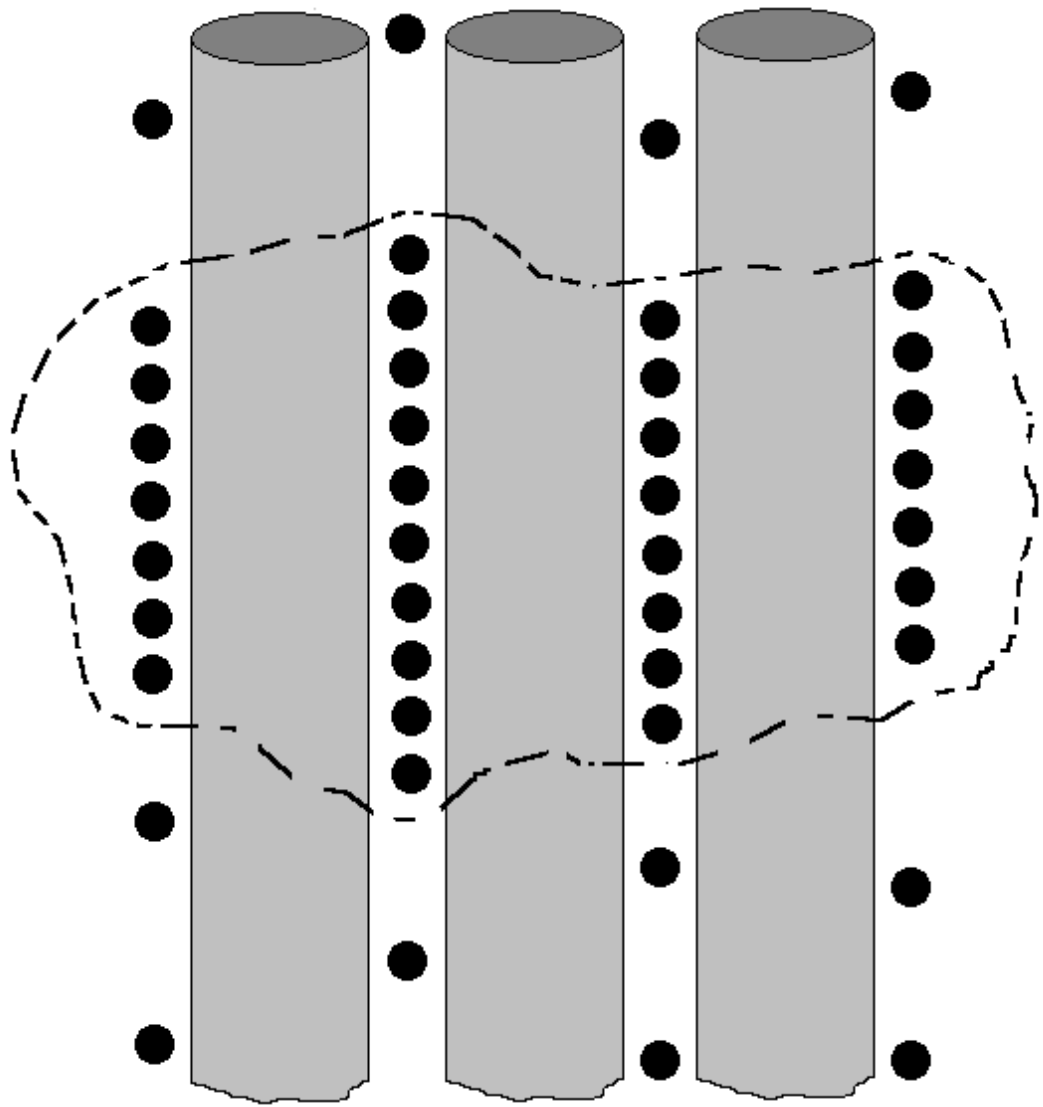
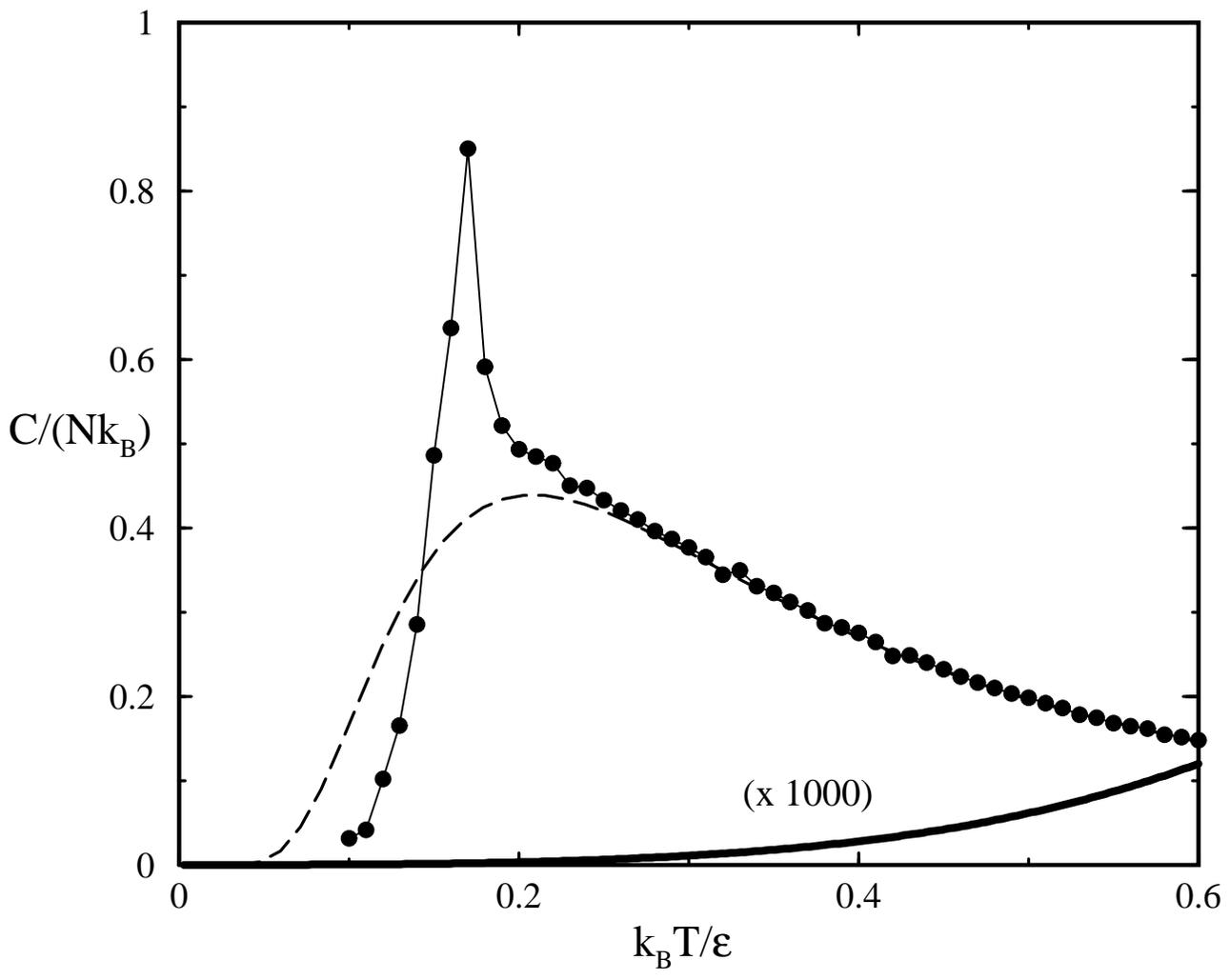
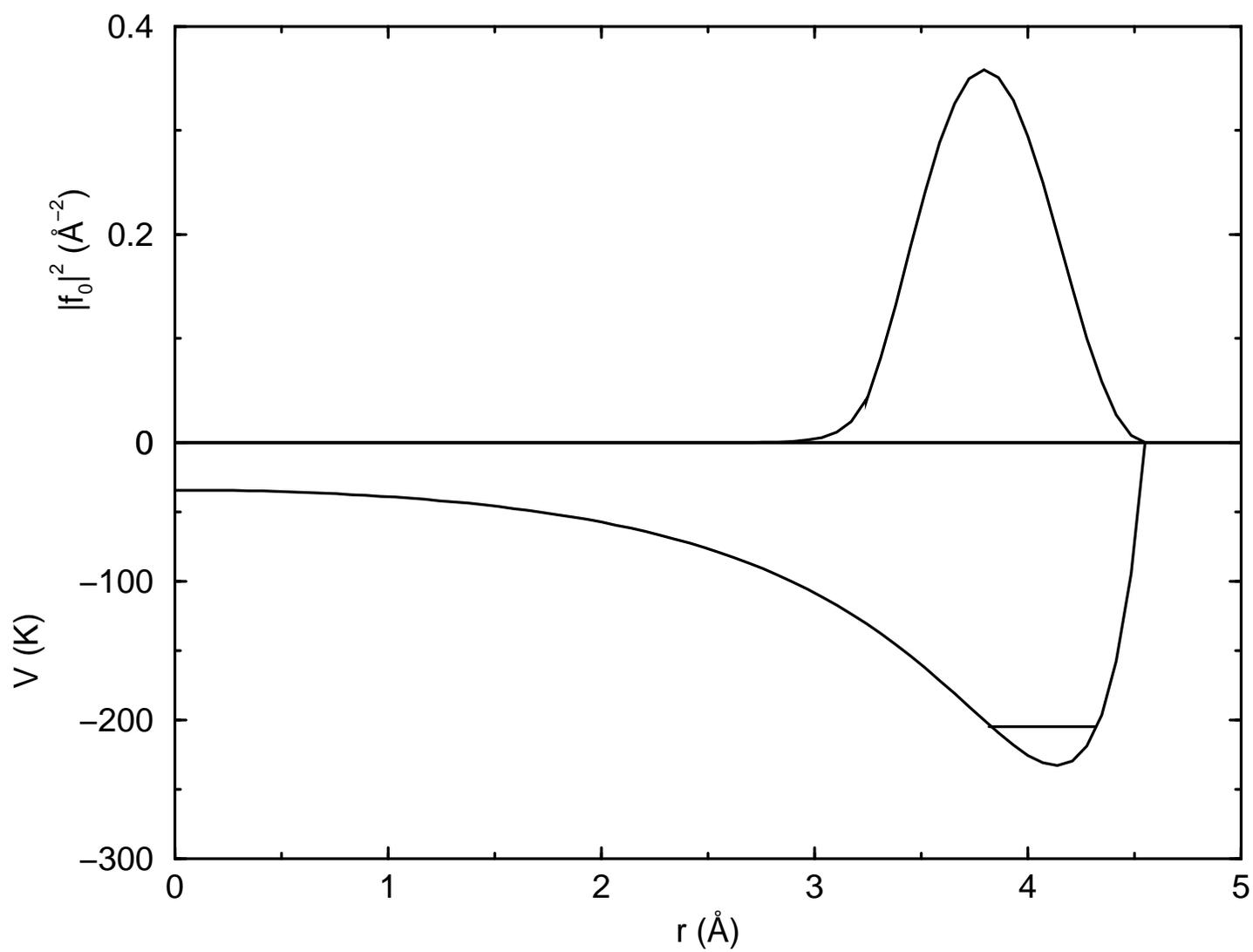
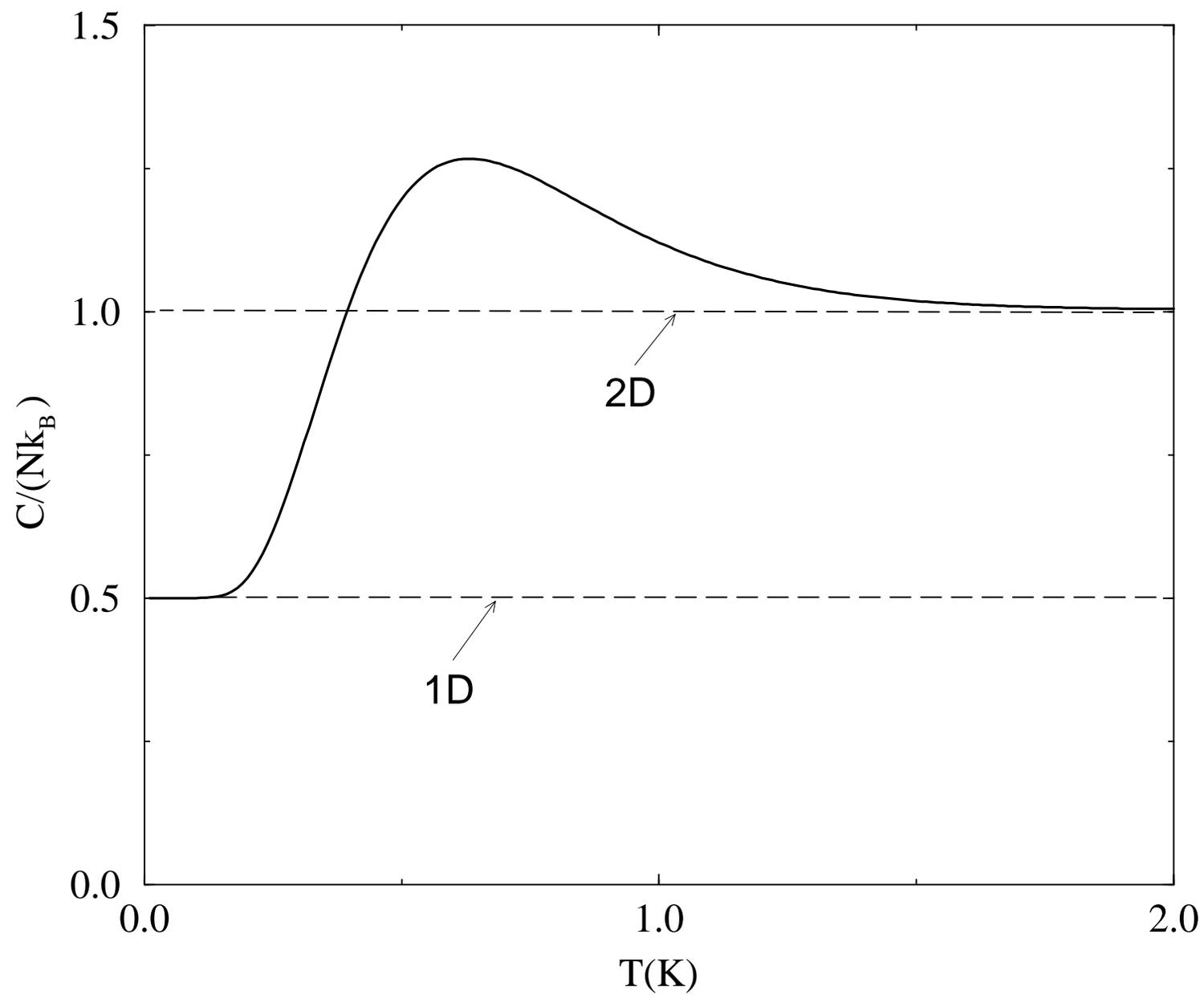
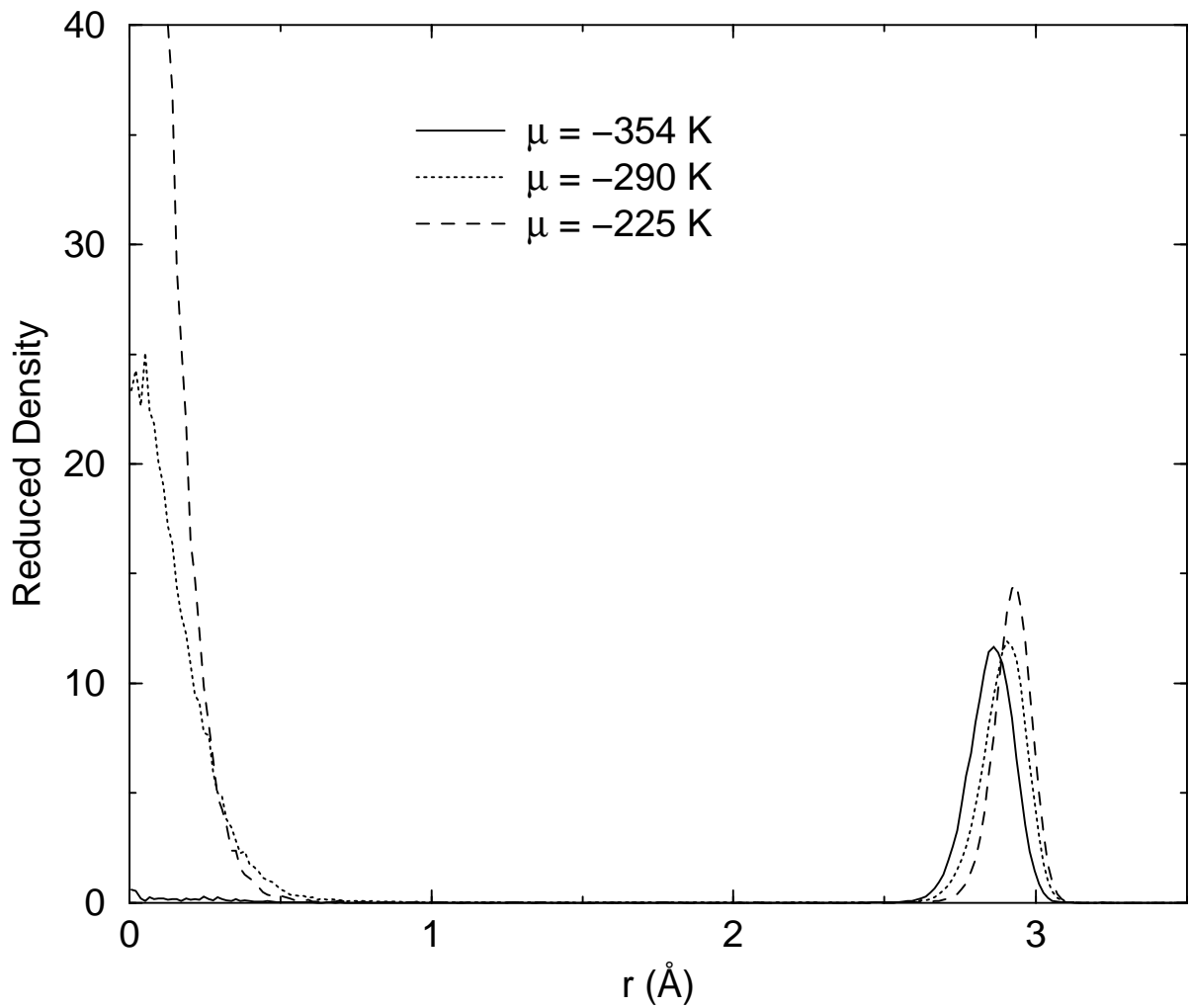
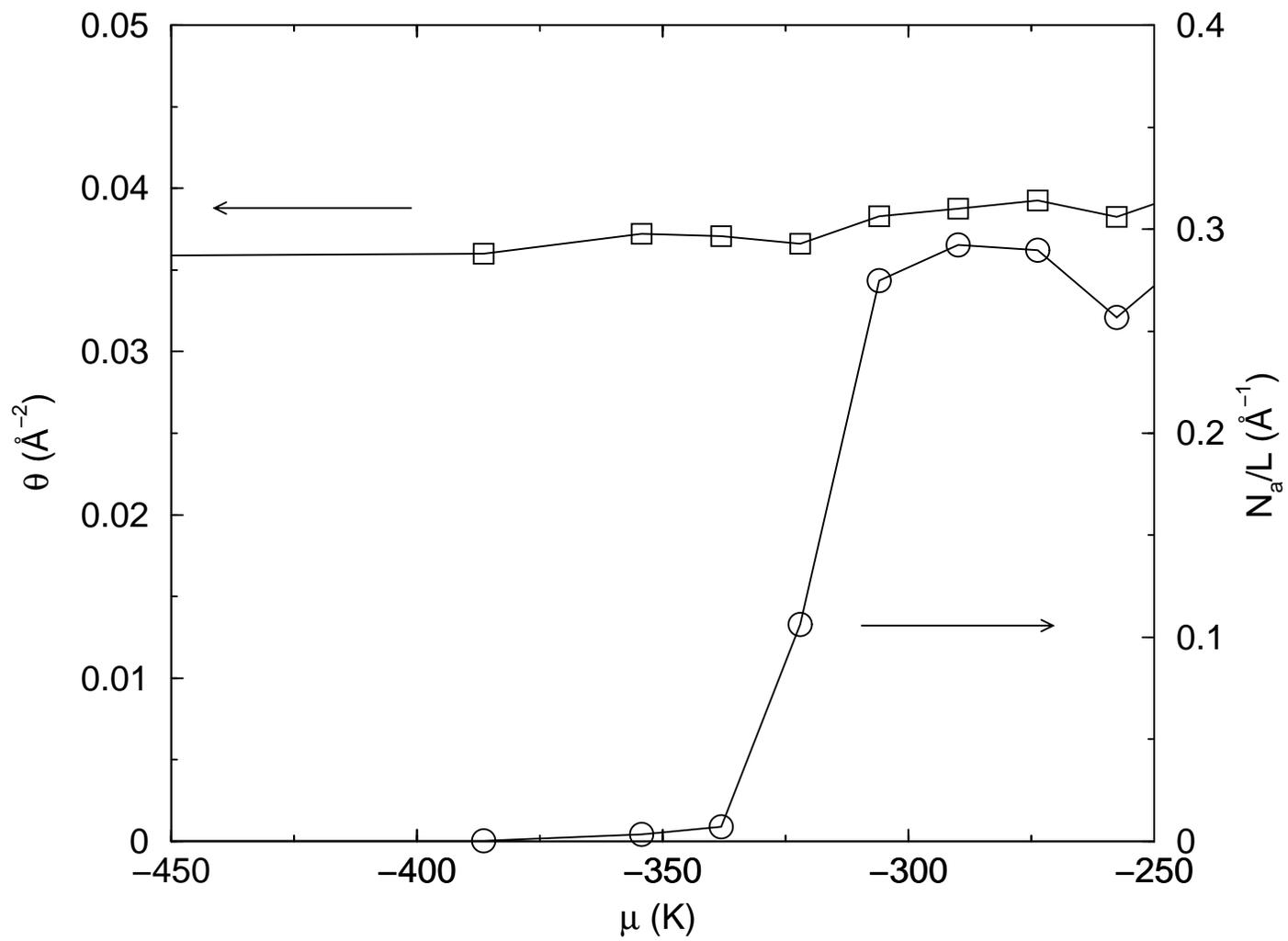
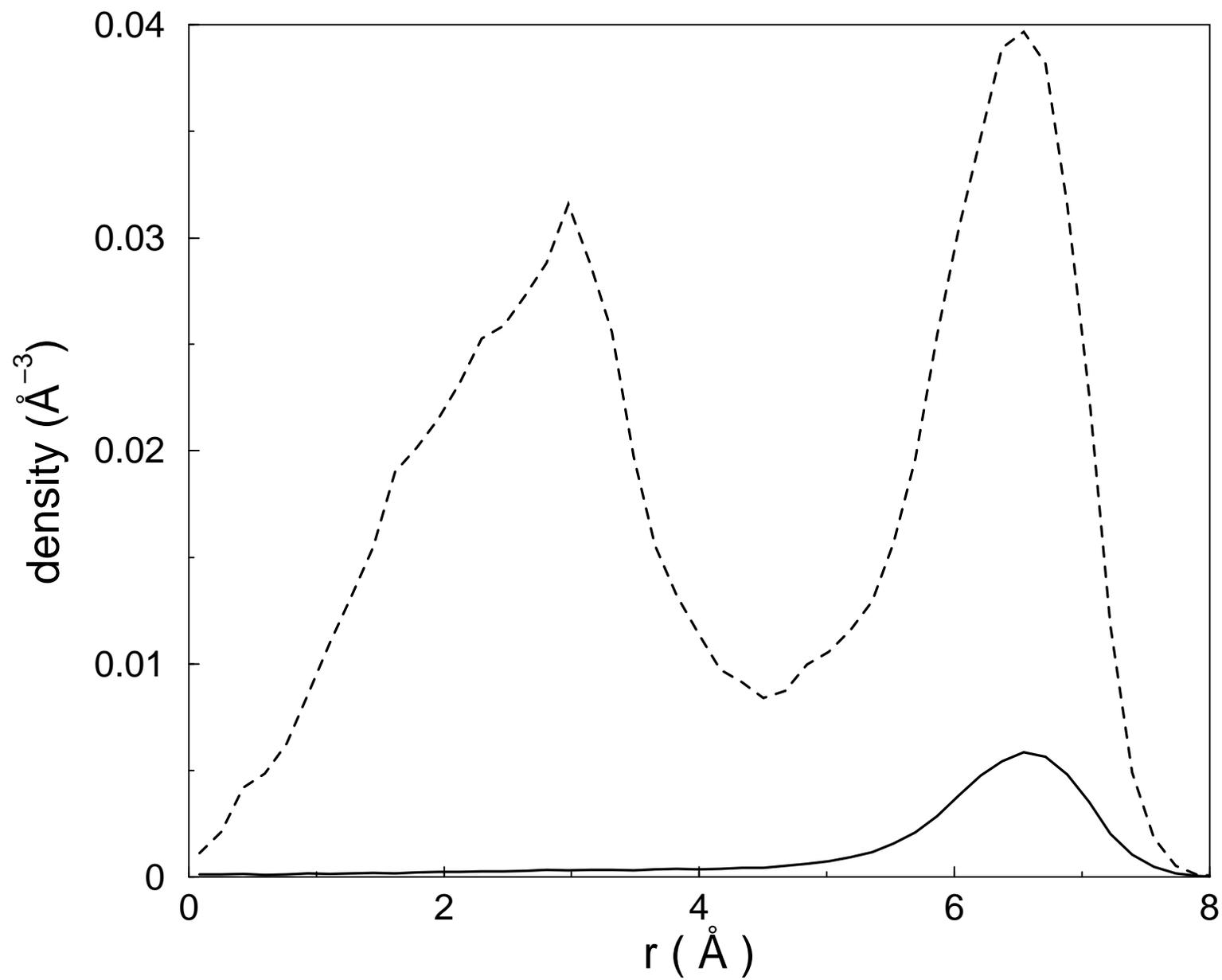
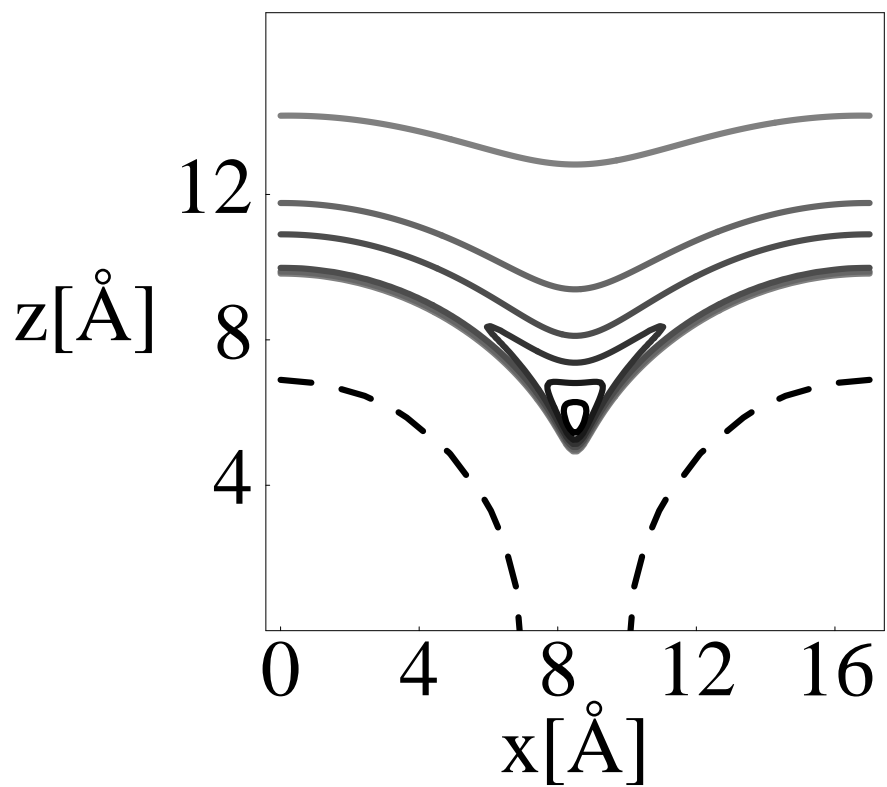
\end{document}